\documentclass{aip-cp}
\pdfoutput=1
\usepackage[numbers,sort&compress]{natbib}
\usepackage{rotating}
\usepackage{graphicx}

\begin{document}

% Title portion
\title{Gamma-Ray Emission from Supernova Remnants\\and Surrounding Molecular Clouds}

\author[aff1]{Stefano Gabici}
\eaddress{gabici@apc.in2p3.fr}
%\author[aff2,aff3]{Author's Name}
%\eaddress{anotherauthor@thisaddress.yyy}

\affil[aff1]{APC, Univ. Paris Diderot, CNRS/IN2P3, CEA/Irfu, Obs. de Paris, Sorbonne Paris Cit\'e, \\75013 Paris, France}
%\affil[aff2]{Additional affiliations should be indicated by superscript numbers 2, 3, etc. as shown above.}
%\affil[aff3]{You would list an author's second affiliation here.}
%\corresp[cor1]{Corresponding author: your@emailaddress.xxx}

\maketitle

\begin{abstract}
Galactic cosmic rays are believed to be accelerated at supernova remnant shocks. Gamma-ray observations of both supernova remnants and associated molecular clouds have been used in several occasions to test (so far quite successfully) this popular hypothesis.
Despite that, a conclusive solution to the problem of cosmic ray origin is still missing, and further observational and theoretical efforts are needed.
In this paper, the current status of these investigations is briefly reviewed.
\end{abstract}

% Head 1
\section{INTRODUCTION}
According to a very popular (but not proven) hypothesis (e.g. \cite{hillas2}), galactic Cosmic Rays (CRs) are accelerated via first order Fermi mechanism at the expanding shocks of SuperNova Remnants (SNRs). Within this framework, SNRs are required to convert $\sim$10\% of the parent supernova explosion energy ($W_{SN} \sim 10^{51}$ erg) into CRs with a broad power law energy spectrum, extending up to rigidities of at least $\sim 3$ PV (which corresponds, for protons, to the position of the knee in the spectrum of CRs observed locally).
Gamma-ray observations of SNRs \cite{dav} or of Molecular Clouds (MCs) associated to SNRs \cite{SNOBs} are considered among the best tools to obtain observational evidence in favour or against this scenario.

The success of the SNR paradigm for the origin of CRs resides mainly in the fact that within this framework: {\it i) }the observed intensity and spectrum of CRs can be reproduced fairly well (e.g. \cite{ptuskinseo}), {\it ii)} the predicted level of anisotropy in the arrival directions of CRs agrees with observations, provided that the CR diffusion coefficient exhibits a mild dependence on energy (close to $D\propto E^{0.3}$) \cite{anisotropies}, and {\it iii)} the predicted CR chemical composition is compatible with the observed one (e.g. \cite{ptuskinseo}).
Moreover, X-ray observations of SNRs revealed the acceleration of electrons up to the multi-TeV domain \cite{koyama}, as well as the presence of narrow filamentary structures (e.g. \cite{ayafilaments,jaccofilaments}) and spatially localised fast time variability \cite{yasrxj}, both interpreted as evidences for the presence of an amplified magnetic field, which is in turn considered a characteristic (though indirect) evidence for hadron acceleration at shocks \cite{bell04}.
Finally, Fermi observations of aged ($\approx 10^4$ yr old) SNRs led to the detection of the characteristic pion-decay signature in their gamma-ray spectrum, proving unambiguously that SNRs indeed accelerate protons with energies at least up to the multi-GeV domain \cite{piondecay}.

On the other hand, some questions remain unanswered, and some pieces of observations are still puzzling.
First of all, no direct observational evidence has been ever reported for the acceleration of PeV protons at SNR shocks.
Also from a theoretical perspective, it is not clear wether SNRs can accelerate protons up to PeV energies at the rate required to explain the CR intensity in the knee region (see \cite{bell13} and discussion below).
Second, a closer look at the chemical composition of CRs suggests that CRs originate from a source material which is a mixture of massive star ejecta and normal interstellar medium \cite{terry}.
This fact brings support to an alternative scenario in which CRs are not accelerated by individual SNRs, but rather at OB stellar associations \cite{SNOBs} or inside the cavities (superbubbles) produced by such stellar associations in the interstellar medium as the result of multiple supernova explosions and stellar winds  \cite{superbubbles}.
Third, data from the Pierre Auger Observatory and the KASCADE-Grande experiment suggest that the transition from galactic to extragalactic CRs takes place at the energy of the ankle, which seems to be beyond the reach of standard diffusive shock acceleration operating at SNR shocks \cite{etienne}.
Finally, to add further uncertainty to the picture, a sharp hardening in the proton and helium CR spectra has been detected by PAMELA \cite{PAMELA} and AMS-02 \cite{ams1,ams2} at a rigidity of few hundreds GV. The origin of these features is still under debate, and might challenge the SNR paradigm, at least in its simplest formulation, which predicts spectra very close to power laws.

\section{COSMIC RAYS FROM SUPERNOVA REMNANTS? GAMMA-RAY BASED TESTS}

An early gamma-ray based test for the origin of CRs at SNRs was proposed in 1994 by Drury, Aharonian and V\"olk \cite{dav}. The test is straightforward: if SNRs are the sources of CRs then each of them should convert about $W_{CR} \sim10^{50}$~erg into CR protons. These CRs would in turn undergo inelastic interactions with the interstellar medium (of density $n \sim$ 0.1...1 cm$^{-3}$) compressed by a factor of $\sim 4$ at the shock, producing neutral pions and thus a gamma-ray flux at the level of:
\begin{equation}
E_{\gamma}^2 F_{\gamma} \sim \frac{W_{CR}}{\ln(E_{max}/E_{min}) ~ \tau_{\pi^0} (4 \pi d^2)} \sim 4 \times 10^{-11} \left( \frac{W_{CR}}{10^{50} {\rm erg}} \right) \left( \frac{n}{{\rm cm}^{-3}} \right) \left( \frac{d}{{\rm kpc}} \right)^{-2} {\rm erg/s/cm}^{2}
\end{equation} 
where $d$ is the distance to the SNR and $\tau_{\pi^0}$ the energy loss time due to neutral pion production. We implicitely assumed an $E^{-2}$ spectrum of accelerated particles extending from $E_{min} \sim 1$ GeV up to $E_{max} =$ few PeV. This flux is well within the reach of Cherenkov telescopes of current generation (H.E.S.S., MAGIC, and VERITAS). For this reason, the detection of several SNRs in TeV gamma rays \cite{felixreview} was welcomed as a success of the SNR hypothesis. However, it is a well known fact that the origin of the detected gamma-ray emission could well be, at least in a number of cases, leptonic (namely, inverse Compton scattering of electrons off cosmic microwave background photons). This means that the detection in gamma rays of SNRs cannot, per se, constitute a proof of the fact that these objects indeed are sources of CRs. An extensive discussion of the leptonic-versus-hadronic dilemma can be found in \cite{piattelli}, where particular emphasis is given to the case of the best studied  gamma-ray bright SNR, RXJ 1713.7-3946.
Despite a large amount of impressive observational results (see e.g. \cite{hessrxj,fermirxj,yasrxj,hessrxjescape}), the hadronic or leptonic origin of the gamma-ray emission detected from this young ($\sim 1600$ yr) SNR is still debated \cite{heinzrxj,giovannirxj,donrxj,finke,zarxj,inoue,merxj}.
Conversely, an hadronic origin of the gamma-ray emission detected from some very young SNRs (e.g. Tycho and Cas A \cite{tycho,casa}, but see also \cite{tychodermer} for a different view) as well as from old SNRs interacting with MCs \cite{piondecay,yascrushed} seems to be preferred over a leptonic one.

\begin{figure}[t]
  \centerline{\includegraphics[width=0.5\textwidth]{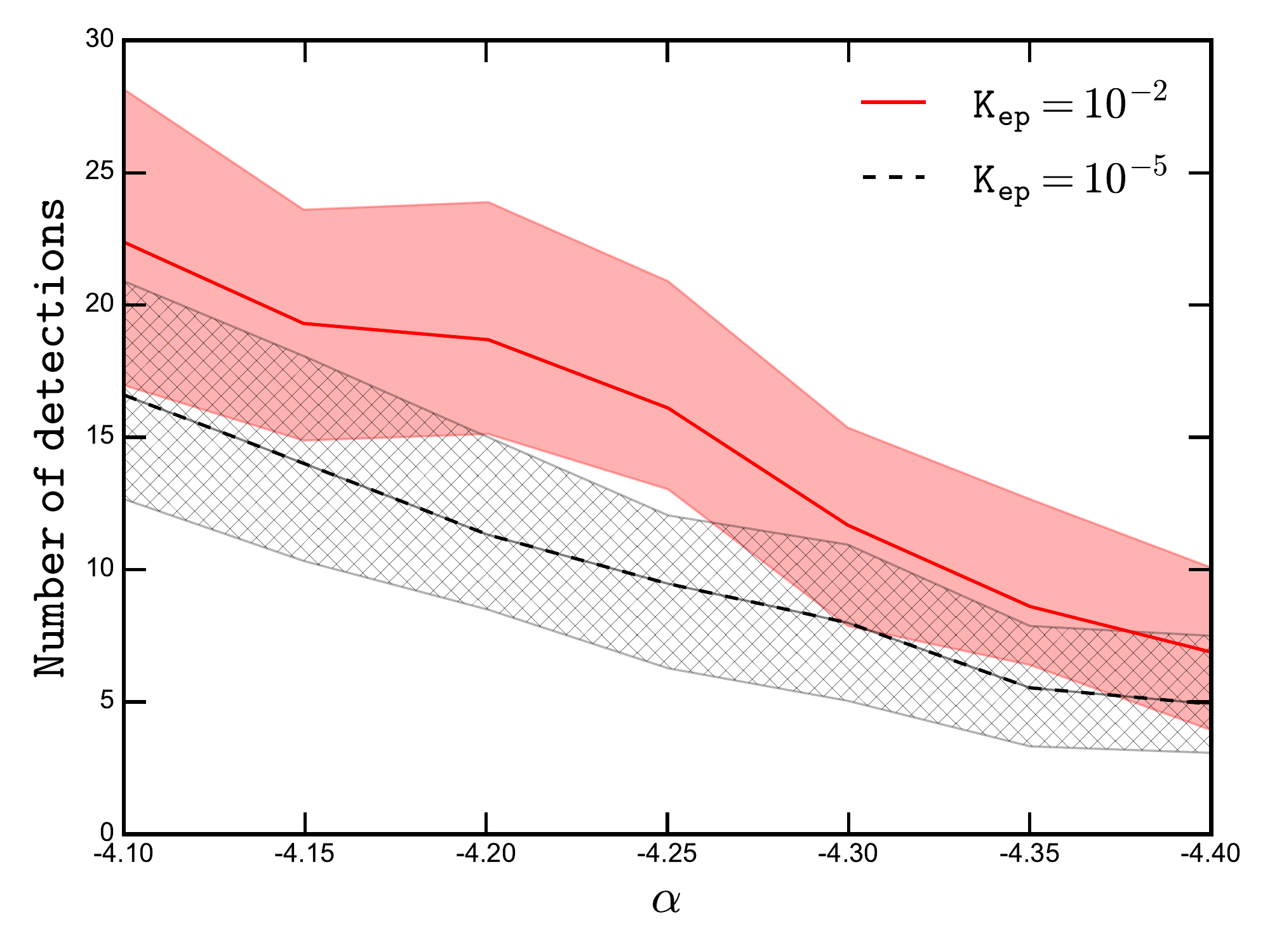}}
\caption{Expected number of SNRs among the TeV sources detected by H.E.S.S. in the galactic region of coordinates $|b| < 3^{\circ}$, $|l| < 40^{\circ}$, with a flux exceeding $\sim 1.5\%$ Crab level ($F(> 1 ~ {\rm TeV}) \sim 3.4 \times 10^{-13}$ ~cm$^{-2}$~s$^{-1}$). The predictions are based on the hypothesis that SNRs are the sources of galactic CRs. The red (black) line refers to a model that includes (does not include) the leptonic contribution to the gamma-ray emission ($K_{ep}$ is the ratio between electrons and protons at injection). Shaded areas show the 1-$\sigma$ fluctuations of the results due to the stochasticity of supernova explosions in the galaxy. The parameter $\alpha$ on the x-axis is the assumed slope of the power law spectrum (in momentum, $\propto p^{-\alpha}$) of CRs accelerated at SNR shocks. Figure adapted from \cite{pierre}.}\label{fig:pierre}
\end{figure}

An additional test has been proposed more recently in \cite{pierre}, where SNRs were not examined on a case-by-case study, but rather as a population of galactic TeV gamma-ray emitters.
The authors of \cite{pierre} computed the number of SNRs which are expected
to be seen in TeV gamma rays above a given flux under the assumption that these
objects indeed are the sources of galactic CRs.
This assumption fixes to about 10\% the typical fraction of the supernova explosion energy which is converted into CRs.
Results were obtained after combining: {\it i)} a Monte Carlo aimed at simulating the location and age of SNR in the galaxy, {\it ii)} a simplified model to describe the acceleration of protons and electrons at SNR shocks, and {\it iii)} the up-to-date knowledge of the large scale atomic and molecular hydrogen distribution in the galactic disk (which impacts on both the SNR dynamical evolution and the production of gamma rays from neutral pion decay).
The outcome of this procedure is shown in Figure \ref{fig:pierre}, where a region of the galactic disk scanned by the H.E.S.S. array of Cherenkov telescope (the H.E.S.S. Galactic Plane Survey, see \cite{donath}) is considered. 
The number of expected detections of SNRs is plotted as a function of the assumed slope $\alpha$ of the differential spectrum (in momentum) of CRs accelerated at SNR shocks.
The red curve refers to a model where the leptonic contribution to the gamma-ray emission (i.e. inverse Compton scattering) is taken into account, while the black curve has been computed by considering the hadronic contribution only.
Within the region under examination, the H.E.S.S. collaboration reported on the firm detection of a handful of SNRs, and of few tens of unidentified (or not firmly identified) sources \cite{donath}. The majority of the firmly identified SNRs exhibit a shell morphology, are of relatively young age (less than $\approx 10^4$ yr), and about half of them are interacting (or suspected to interact) with a MC \cite{donath}. It seems, then, that expectations are consistent with observations, as long has the slope of the spectrum of accelerated CRs is steeper than $E^{-2}$ (or $p^{-4}$), and harder than $\approx E^{-2.4}$ (or $p^{-4.4}$). 
A harder (steeper) spectrum would result into too many (too few) detections.
It is interesting to note that very similar constraints on the slope of the injection spectrum of CRs are obtained from studies of the CR propagation in the galaxy (see e.g. \cite{strongreview}), and that source spectra steeper than $E^{-2}$ are also predicted by the non-linear theory of shock acceleration (e.g. \cite{damiano}).

Finally, the Fermi collaboration recently released the first catalogue of GeV SNRs \cite{fermisnrcatalogue}. It consists of 30 sources most likely associated with SNRs, 14 marginal associations and 245 flux upper limits. For 25 of such SNRs, an estimate of both the upstream gas density and of the SNR distance can be found in the literature. For many others, only the distance is known, and a typical upstream density of 1 cm$^{-1}$ can be considered a quite reasonable guess. These two quantities, together with the gamma-ray flux (or flux upper limit), can be used to constrain the CR content of a given SNRs.
If one further assumes that the gamma-ray emission is entirely due to the decay of neutral pions, then an upper limit on the energy in form of hadronic CRs, $W_{CR}$, can be obtained.
The upper limits obtained in this way span several orders of magnitude, going from several $10^{49}$ erg up to several $10^{52}$ erg, which exceeds by far the supernova explosion energy $W_{SN} \sim 10^{51} {\rm erg}$.
Such a broad range of values does not allow to obtain a very stringent constrain on hadronic CR production in SNRs, but nevertheless it definitely does not contradict the hypothesis that SNRs should convert a fraction $\epsilon_{CR} = W_{CR}/W_{SN} \approx 0.1$ of the explosion energy, i.e. $W_{CR} \approx 10^{50}$ erg, into accelerated protons, in order to be the sources of galactic CRs. 

Of particular interest for GeV observations are the SNRs interacting with massive MCs, because the spatial correlation between gamma rays and gas density points towards an hadronic origin of the emission \cite{yascrushed}. For several of such systems the value of $\epsilon_{CR}$ inferred from Fermi data largely exceeds unity, which is clearly an unphysical result. In fact, this happens because in this case the density experienced by CR protons in the MC is much larger than the typical upstream density found in the literature, and this artificially enhances the derived value of $\epsilon_{CR}$.
The large number of SNRs found by Fermi in the $\epsilon_{CR} > 1$ region suggests that associations with MCs are quite common (which is not a surprise \cite{SNOBs}) and thus that most of the emission is hadronic.
Indeed, a simplified calculation shows that the number of SNRs detected by Fermi is consistent with the canonical value $\epsilon_{CR} \sim 0.1$ only for values of the density experienced by CRs of the order of tens of particles per cubic centimetre.
This indicates that most of the SNRs detected by Fermi (except for the known young and isolated ones) are SNR/MC interacting systems. 

\section{ARE SUPERNOVA REMNANTS PROTON PEVATRONS?}
Though the three gamma-ray based tests briefly described in the previous Section bring significant support to the SNR hypothesis, they cannot be considered as proofs for it.
One of their main limitations is that, being based on GeV and TeV observations, they cannot provide insights on the acceleration of protons up to the knee region. In other words, they cannot tell us wether SNRs are proton PeVatrons or not.

In order to prove observationally that SNRs accelerate PeV particles one would need either to detect multi-TeV neutrinos or to measure gamma-ray spectra extending without significant attenuation up to the multi-TeV domain. 
This is because the inelastic interactions of CR protons of energy $E_p$ with the interstellar gas lead to the production of neutrinos of energy $E_{\nu} \sim 50 ~(E_p/1~{\rm PeV})$ TeV and gamma rays of energy $E_{\gamma} \sim 100 ~(E_p/1~{\rm PeV})$ TeV \cite{kelner}. 
Moreover, the most relevant competing mechanism of production of gamma rays, inverse Compton scattering, is inefficient at such large energies, due to the suppression of the cross section in the Klein-Nishina regime \cite{felixbook}. 
Unfortunately, none of these evidences has ever been found for a SNR, and for this reason the hunt for PeVatrons constitutes one of the major goals of future gamma-ray facilities such as the Cherenkov Telecope Array \cite{cta}.
Finally, to make the picture even more intriguing, the H.E.S.S. Collaboration recently reported on the first discovery of a galactic PeVatron, most likely associated with the supermassive black hole located at the galactic centre \cite{pevatron}. Thus, we now know that proton PeVatrons other than SNRs do exist in the galaxy.

In general, an astrophysical object is considered capable of accelerating particles up to a given energy $E_{max}$ if it satisfies a condition known as the {\it Hillas criterion}, i.e., the object must be able to confine particles of energy $E_{max}$ \cite{hillas1,hillas2}.
For SNRs this is equivalent to say that the diffusion length of particles ahead of the shock should not exceed a given fraction $\xi \sim 0.05$ of the SNR shock radius: $D(E_{max})/u_s = \xi R_s$, where $D(E_{max})$ is the Bohm diffusion coefficient for protons of energy $E_{max}$, and $u_s$ and $R_s$ are the SNR shock speed and radius, respectively \cite{pz2005}.
In numbers:
\begin{equation}
\label{eq:hillas}
E_{max} \sim \frac{1}{3} \left( \frac{R_s}{\rm pc} \right) \left( \frac{u_s}{1000~{\rm km/s}} \right) \left( \frac{B}{\mu {\rm G}} \right) ~ \rm TeV
\end{equation}
which implies that for a typical interstellar magnetic field $B \sim 3 ~\mu$G, and during the ejecta dominated phase of the SNR evolution, characterised by $u_s \approx 10000$ km/s and $R_s$ of a few pc, the maximum energy falls short of the knee by a large factor.

This fact was first remarked by Lagage and Cesarsky in 1983 \cite{lagage} and remained for about 20 years one of the main problems of the SNR paradigm for the origin of CRs.
It is clear from Equation \ref{eq:hillas} that the only way to reach PeV energies at a SNR shock is to assume that some mechanism of magnetic field amplification operates there.
In 2004, Bell suggested that a CR induced plasma instability might indeed be responsible for an amplification of the field by a factor of several tens or even hundreds, making the acceleration of PeV particles possible \cite{bell04}.
However, in a more recent study, Bell et al. \cite{bell13} pointed out that the acceleration up to PeV energies requires extremely large shock speeds, that can be attained only during the very early phase (at most few tens of years) of a SNR evolution.
This scenario for PeV particle acceleration at SNR shocks, if confirmed, has two extremely important implications:
first of all, an estimate of the number of SNRs which are currently accelerating PeV particles can be written as $\sim 1 (\Delta t_{\rm PeV}/30~{\rm yr}) (30 ~ {\rm yr } \times \nu_{SN})$, where $\Delta t_{\rm PeV}$ is the duration of the PeV particles acceleration phase in SNRs, and $\nu_{SN} \sim (30 ~ {\rm yr})^{-1}$ is the typical supernova explosion rate in the galaxy.
Thus, due to the very short duration of the PeV phase, the number of SNRs active as PeVatrons in the galaxy is {\it at most} very few, posing obvious observational problems.
Second, the question arises whether in such a short time SNRs can convert a sufficient amount of energy into PeV protons, and explain the observed intensity of CRs in the region of the knee.
This is because a large fraction of the available energy is dissipated at SNR shocks at the transition between the ejecta dominated and the Sedov phase, which happens when the SNR has a typical age of few centuries, which is significantly larger then $\Delta t_{\rm PeV}$.

A possible way-out is to invoke a hypothetical, more efficient mechanism for magnetic field amplification at shocks. Indeed, scenarios alternative or complementary to that proposed by Bell do exist. Of particular interest is a model developed by Drury and collaborators, where the amplification of the turbulent magnetic field is driven by the presence of a CR pressure gradient upstream of the shock \cite{lukeB}. Under certain circumstances a significant fraction of the shock ram pressure might be converted in this way into turbulent magnetic pressure, while this fraction is smaller (at the percent level) for the Bell mechanism \cite{bell13}.

\section{STUDYING COSMIC RAY ESCAPE WITH MOLECULAR CLOUDS}
The question about the maximum energy particles can attain at SNR shocks is intimately connected to the way in which particles escape SNR shells. Due to its highly non-linear nature (CRs are believed to trap themselves within SNR shells \cite{bell13}), this is one of the less understood aspects of diffusive shock acceleration. Insights on the mechanism of escape can be obtained from gamma-ray observations because runaway CRs can interact with massive MCs located in the vicinity of the parent SNR and make the former bright gamma-ray sources \cite{atoyan,me2009}. To date, a clear detection of gamma rays from the vicinity of SNRs has been reported by the Fermi collaboration for the SNR W44 \cite{W44fermi}, and by the H.E.S.S. and Fermi collaborations for the SNR W28 (see left panel of Figure \ref{fig:mev}) \cite{W28hess,W28fermi}. Other SNR/MC associations might also be interpreted as the result of the escape of CRs and their interaction in the dense gas of the cloud (e.g. \cite{diego,ohira,gerd}).
Moreover, very recently the H.E.S.S. collaboration reported on the detection of gamma-ray emission beyond the X-ray shell of the SNR RX J1713.7-3946, which can be interpreted either as a signature of particle escape, or as the first detection of a shock CR precursor \cite{hessrxjescape}.

A toy model aimed at describing the propagation of runaway CRs around SNRs can be developed under the assumption of an homogeneous and isotropic CR diffusion coefficient. At any given time, the energy spectrum of CR protons escaping a SNRs is believed to be a function very peaked around $E_{p}(t_{age})$, which is in turn a (not very well known) decreasing function of the SNR age $t_{age}$ (i.e. higher energy particles escape earlier than lower energy ones) \cite{pz2005}. A time $\Delta t = t_{age}-t_{esc}$ after escape, CR protons of energy $E_{p}(t_{esc})$ propagate away from the SNR shock a distance $R_d \sim \sqrt{6 ~D~ \Delta t}$. If sufficiently old SNRs, and CR of sufficiently large energy are considered, then $t_{esc}(E_{p}) \ll t_{age}$, $\Delta t \sim t_{age}$, and $R_s \ll R_d$, and the intensity of CRs of energy $E_p$ within a distance $R_d$ from the SNR centre scales as $f_{\alpha} W_{CR}/R_d^3 \sim f_{\alpha} \epsilon_{CR} W_{SN}/R_d^3$, where $\epsilon_{CR}$ is a CR acceleration efficiency and $f_{\alpha}$ a function that takes into account the shape of the injection spectrum of runaway CRs.
If a MC of mass $M_{cl}$ is located within a distance $R_d$ from the SNR, then gamma rays of energy $E_{\gamma} \sim 0.1 \times E_{p}$ will be produced as the result of the hadronic interactions between runaway CRs and the dense gas of the cloud, at a flux level proportional to:
\begin{equation}
\label{eq:simple}
F_{\gamma} \propto \frac{f_{\alpha} \epsilon_{CR} W_{SN} M_{cl}}{R_d^3~ d^2}
\end{equation}
where $d$ is the distance to the observer.
This simple model can be used to show that gamma-ray observations of SNR/MC associations can be used to tackle at least two crucial problems in CR astrophysics, briefly described below.

{\it 1) Search for PeVatrons --} A clear and unambiguous signature for the presence of PeV protons is the detection of gamma ray spectra extending up to energies of $\sim 100~ (E_p/{\rm PeV})$ TeV. As stated in the previous Section, determining observationally whether or not SNRs are proton PeVatrons is challenging. This is not only because the multi-TeV domain region of the electromagnetic spectrum is, to date, not very well explored, but also because the duration of the PeV-phase, $\Delta t_{\rm PeV}$, in the lifetime of a SNR is believed to be very short, of the order of tens of years. After this time, PeV particles escape the SNR shock, but their presence can still be revealed by means of the radiation they produce in the medium surrounding the parent SNR \cite{me2007}. 
If a MC is located at a distance $R$ from the SNR, the duration of the multi-TeV gamma-ray emission produced by runaway CRs into the cloud is expected to last $\Delta t_{\gamma} \sim R^2/6 ~D_{\rm PeV}$, which is the characteristic propagation time of PeV particle over a distance $R$. Here, $D_{\rm PeV}$ is the diffusion coefficient of PeV CRs.
In order to have $\Delta t_{\gamma} > \Delta t_{\rm PeV}$ and then enhance the probability to reveal the interactions of PeV protons a condition on the diffusion coefficient has to be satisfied, namely, $D_{\rm PeV} < R^2/6 ~ \Delta t_{\rm PeV} \sim 10^{31} (R/100~{\rm pc})^2 (\Delta t_{\rm PeV}/30~{\rm yr})^{-1}$ cm$^2$/s, which is a quite plausible number.

Even though the acceleration of PeV protons in SNRs (or their presence in the vicinity of SNRs after their escape) has never been observed, it is remarkable that the first PeVatron discovered in our galaxy, most likely associated with the central supermassive black hole, has been discovered by using this very same technique, i.e., by observing the diffuse gamma-ray emission produced by the interaction of CRs escaping from the galactic centre with the dense gas in the central molecular zone \cite{pevatron}.
Future facilities such as the Cherenkov Telescope Array are expected to contribute significantly to the indirect search of PeVatrons, having the sensitivity needed to detect the multi-TeV emission generated by runaway CRs in the vicinity of the parent SNRs \cite{sabrinarxj}.

\begin{figure}[t]
  \centering
  \includegraphics[width=0.55\textwidth]{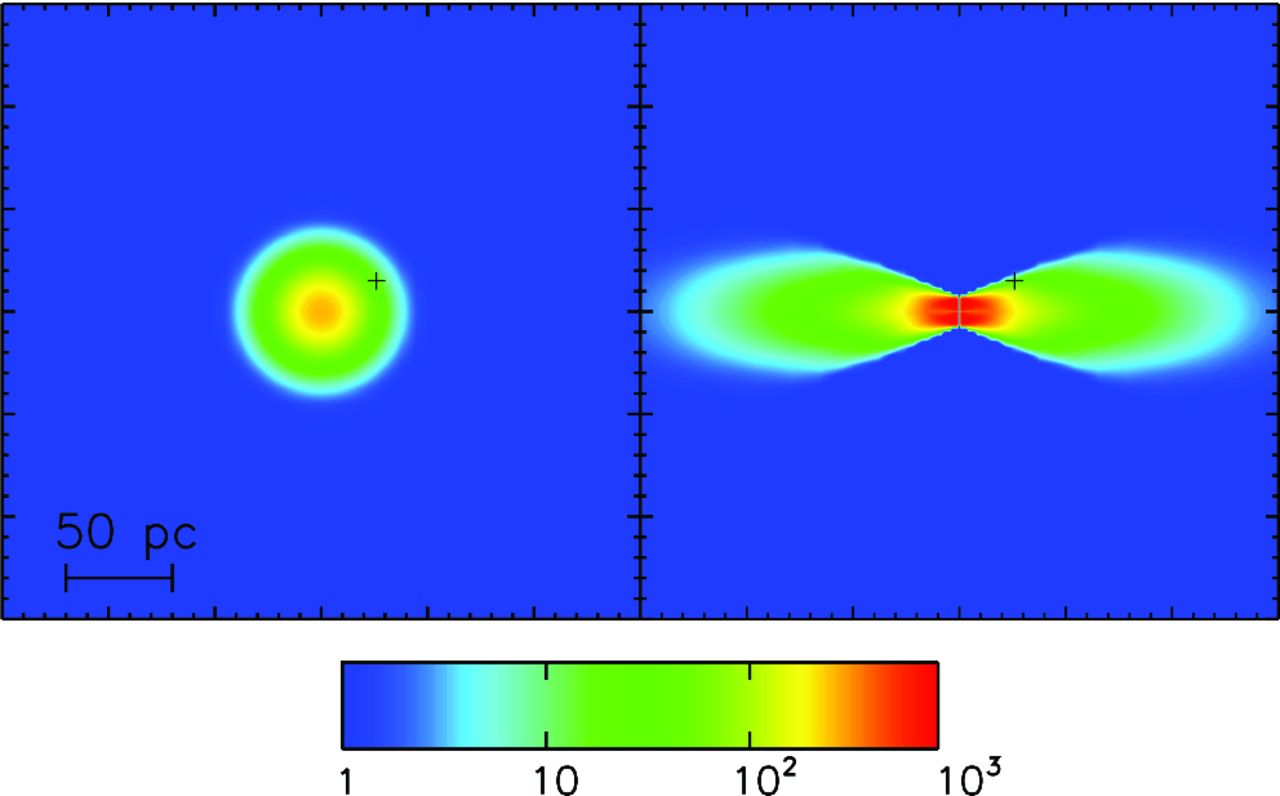}
  \includegraphics[width=0.45\textwidth]{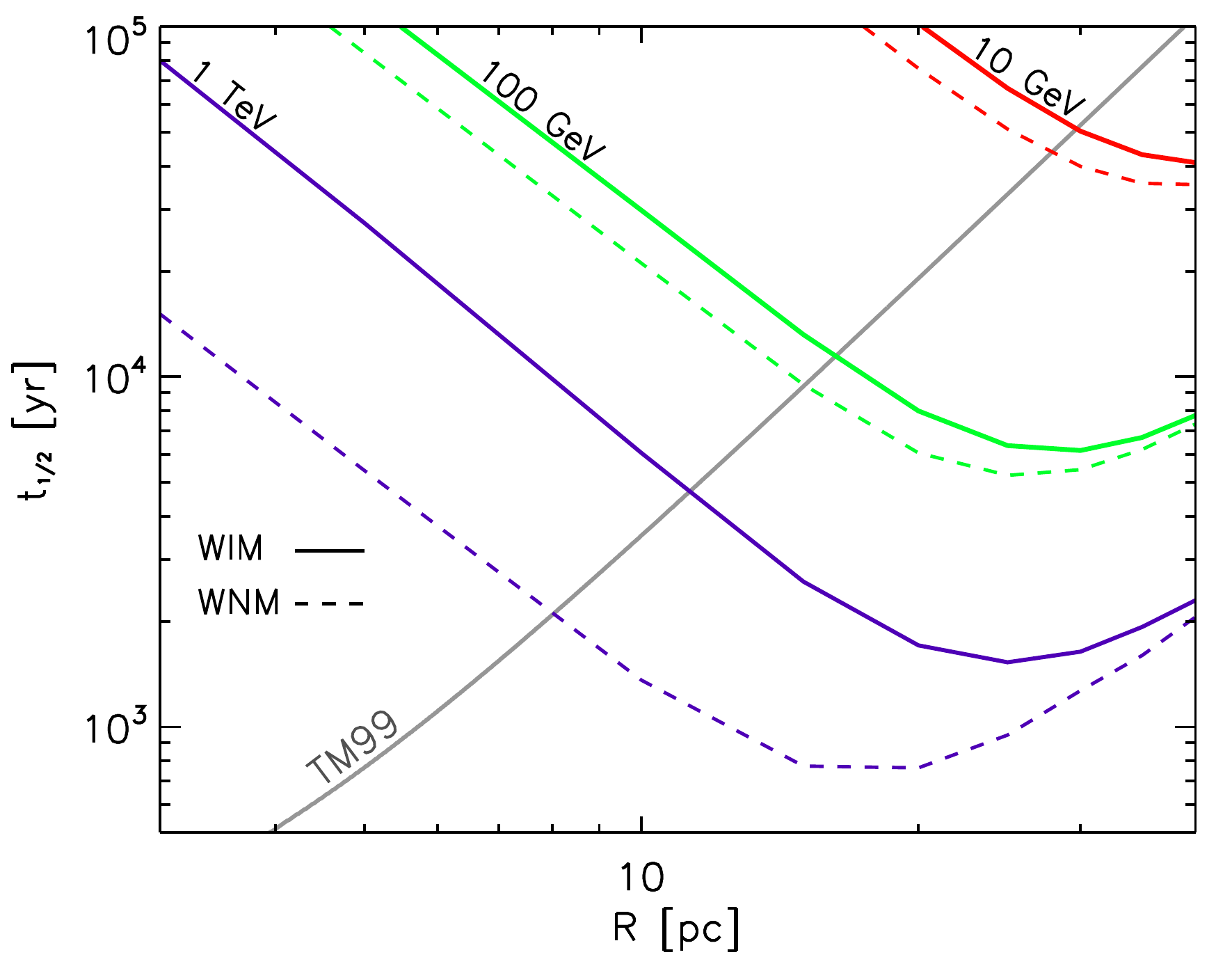}
\caption{{\bf Left:} CR overdensity around a typical SNR (explosion energy $W_{SN} = 10^{51}$ erg, mass of the supernova ejecta $1.4 M_{\odot}$, interstellar gas density $1$ cm$^{-3}$, CR acceleration efficiency $\epsilon_{CR} = 0.1$) for a particle energy of 1 TeV at a time 10 kyr after the explosion. The left-hand map refers to an isotropic diffusion coefficient of CRs equal to $D = 5 \times 10^{26} (E/10~ {\rm GeV})^{0.5}$ cm$^2$/s, while the right-hand map refers to an anisotropic diffusion scenario where CR propagate along the magnetic field lines with a diffusion coefficient $D_{\parallel} = 10^{28} (E/10~ {\rm GeV})^{0.5}$ cm$^2$/s. The transport perpendicular to the mean field direction is assumed to be due only to the wandering of magnetic field lines, with a diffusion coefficient of $D_m = 1$ pc, and $(\delta B/B)^2 = 0.2$ (see \cite{lara1} for more details). The black cross marks a position at which the CR overdensity is equal in the two panels, i.e., a MC located there would emit the same gamma-ray flux in both scenarios \cite{lara1}. {\bf Right:} Half-time of the CR cloud (see the text) as a function of its initial radius $R$. Red, green, and purple lines refer to particle energies of 10, 100, and 1000 GeV. Solid and dashed lines refer to the WIM and WNM phases of the interstellar medium. The total energy of CRs is set to $W_{CR} = 10^{50}$ erg and their spectrum at injection is proportional to $E^{-2.2}$. The black solid line represents the relationship between SNR radius and age according to \cite{truelove}. Figure from \cite{lara2}.}
\label{fig:clouds}
\end{figure}

{\it 2) Constrain the CR diffusion coefficient --} The CR diffusion coefficient is a very poorly known physical quantity. Its typical value in the galaxy is of the order of $\approx 10^{28}-10^{29}$ cm$^2$/s for a proton energy of $\sim 10$ GeV, and it scales with particle energy as $E_p^{\delta}$ with $\delta \sim 0.3 ... 0.5$ \cite{strongreview}. This value has to be considered as a time and space average, and very little is known about the diffusion coefficient at specific locations in the galaxy.
Equation \ref{eq:simple} shows that an estimate of the diffusion coefficient can be obtained from gamma-ray observations of MCs located in the vicinity of SNRs if the detected gamma-ray emission can be confidently ascribed to the interaction of runaway CRs coming from the nearby SNR.
The gamma-ray flux and spectrum of the MC provide the values of $F_{\gamma}$ and $f_{\alpha}$, while low energy observations (mainly the detection of CO molecular lines in the radio domain) of the MC give the values of $M_{cl}$ and $d$. These measurements, combined with the fact that the supernova explosion energy has a quite standard value of $W_{SN} \sim 10^{51}$ erg, and to the obvious condition $\epsilon_{CR} < 1$ ($\epsilon_{CR} \sim 0.1$ if SNRs are the sources of CRs), impose a constrain on the diffusion length $R_d \sim \sqrt{6 ~ D ~ t_{age}}$.
If also the age of the SNR is known a constrain on the CR diffusion coefficient can be finally derived.

The SNR W28, surrounded by gamma-ray bright MCs (see left panel of Figure \ref{fig:mev}) constitutes an ideal system for this kind of studies.
In \cite{sf2a} a diffusion coefficient a factor of 10...100 smaller than the typical galactic one was derived for $\sim$ TeV particles, based on gamma-ray observations.
Similar results have been obtained from studies of a number of other SNR/MC associations, suggesting that a suppression of the CR diffusion coefficient in the vicinity of CR sources might be a common fact \cite{ohira,gerd}.
However, such a result was obtained after assuming an isotropic CR diffusion coefficient.
In fact, the magnetic field in the galaxy is known to be correlated over scales of $\sim 50-100$ pc, and the assumption of isotropic diffusion of CRs seems difficult to be justified over spatial scales smaller than that.
CRs are believed to propagate mainly along magnetic field lines, and this fact significantly affects the interpretation of gamma-ray observations of SNR/MC associations \cite{lara1}.
This is mainly due to the fact that the volume occupied by runaway CRs of a certain energy and at a given time after their escape from the SNR no longer scales as $R_d^3$, but rather as $\sim R_d \times R_s^2$, where $R_s$ is the radius of the SNR at the time of the particle escape.
It is then straightforward to see that the constrain on the {\it parallel} (along the field lines) diffusion coefficient obtained from gamma-ray observations would be in this case less stringent by a factor of $\sim (R_d/R_s)^{4/3} \approx 22~(R_d/100~{\rm pc})^{4/3} (R_s/10~{\rm pc})^{-4/3}$ \cite{lara1}.
This factor of the order of few tens quantifies our ignorance in the determination of the CR diffusion coefficient, i.e., it is the difference in the measurement of $D$ if the two most extreme assumptions are made of a fully isotropic or fully anisotropic diffusion.
The difference between these two extreme scenarios is illustrated in the left panel of Figure \ref{fig:clouds}.

To complement the observational/phenomenological studies described so far, theoretical investigations aimed at computing the CR diffusion coefficient in the vicinity of CR sources have been presented in several publications.
Pioneering attempts in this direction dates back to the seventies \cite{skilling,cesarsky75}, and recent ones include references \cite{plesser,lara2,dangelo}.
All these approaches consider a magnetic flux tube containing the SNR shock, which acts as the particle injector. 
The transport of CRs along the flux tube is assumed to be regulated by the resonant scattering off Alfv\'en waves, i.e. CRs resonate with waves of wave number $k = 1/r_L(E_p)$, where $k$ is the wave number and $r_L$ is the Larmor radius of CR protons of energy $E_p$. 
The (normalized) energy density $I(k)$ of Alfv\'en waves is defined as: $(\delta B/B)^2 = \int {\rm d} \ln k
~I(k)$ where $B$ is the ambient magnetic field and $\delta B$ the amplitude of the magnetic field fluctuations.
If $\delta B \ll B$ we are in the quasi-linear regime and the CR diffusion coefficient along the flux tube is given by the ratio between the Bohm diffusion coefficient and the energy density of resonant waves: $D(E_p) = D_B(E_p)/I(k)$.
Diffusion across field lines can be neglected and the problem is one-dimensional.
The problem becomes non-linear when one considers the fact that CRs streaming along the flux tube at a velocity larger than $v_A$ induce a growth of pre-existing Alfv\'en waves. The time evolution of waves and CRs along the flux tube (oriented along the z-axis) are then described by a system of coupled partial differential equations:
\begin{equation}
\label{eq:system}
\frac{\partial P_{CR}}{\partial t} = \frac{\partial}{\partial z} \left( \frac{D_B}{I} \frac{\partial P_{CR}}{\partial z} \right) ~~~ ; ~~~ \frac{\partial I}{\partial t} = -v_A \frac{\partial P_{CR}}{\partial z} - 2 \Gamma_d I + Q
\end{equation}
where $P_{CR}$ is the pressure of resonant CRs in units of the mgnetic field energy density $B^2/8 \pi$, the term $-v_A \partial P_{CR}/ \partial z$ represents the growth of waves due to CR streaming (it can be interpreted as the rate of work done by the CRs in scattering off the waves), $\Gamma_d$ is a damping term for waves and $Q$ accounts for possible contributions to the Alfv\'enic turbulence other than CRs.
The choice of the damping term is a crucial one. A large fraction of the volume of the galactic disk is occupied by the Warm Ionised Medium (WIM) and Warm Neutral Medium (WNM), characterised by a ionization fraction of 0.9 and 0.02, respectively. In both cases, the number density of neutrals is large enough to make ion-neutral friction the dominant damping mechanism for Alfv\'en waves \cite{zweibel}.

We can now solve the system of Equations \ref{eq:system} by assuming that CRs of a given energy E are initially uniformly distributed over a region (a cloud) of radius R. Following \cite{malkov,lara2}, we define the half-time of the CR cloud as the time at which half of the CRs which were initially into the cloud have diffusively escaped outside of its initial boundary. 
The half-time is plotted in the right panel of Figure \ref{fig:clouds} as a function of the initial cloud radius R.
Here, we propose to consider the half-time of the CR cloud as a rough estimate of the escape time of CRs from the region of initial size $R$. Though we are aware of the fact that the process of particle escape from SNRs is still poorly understood, we propose to extend this operational definition of escape time to SNRs also.
For this reason, we plot as a black line in Figure \ref{fig:clouds} the relationship
between the age and the radius of a SNR expanding in a homogeneous medium \cite{truelove}.
The intersections between the black and the colored lines in the right panel of Figure \ref{fig:clouds} indicates at which time CRs of a given energy leave the SNR. Within this framework, we can reproduce the qualitative result that higher energy particles escape SNRs earlier than lower energy ones.
Nava et al. \cite{lara2} found out that the half-time also gives a rough estimate of the time interval during
which CR streaming instability amplifies the Alfv\'enic turbulence in the surrounding of SNRs. After estimating the half-time of the cloud for typical SNR parameters, they found it to be a decreasing function of particle energy. As a consequence, the propagation of $\sim 10$ GeV CRs is affected
for several tens of kiloyears, while for $\sim 1$ TeV CRs, such time-scale reduces to few kiloyears.

The relatively meagre number of SNR/MC associations detected in gamma-rays makes it difficult to properly constrain these theoretical models. However, it is beyond any doubt that observations performed by instruments of next generation (such as the Cherenkov Telescope Array) will increase significantly the statistics of detections and provide meaningful constraints to models.

\section{PENETRATION OF COSMIC RAYS IN MOLECULAR CLOUDS}
An implicit assumption adopted so far is that of unimpeded penetration of CRs into MCs. 
In fact, such an assumption needs to be justified, because the exclusion of CRs from MCs might have an impact on the gamma-ray emission from clouds and on their level of ionisation (an issue that will be discussed in the next Section).
The most straightforward way to test the level of penetration of CRs into MCs is to perform gamma-ray observations of clouds in the GeV domain. At these energies, the emission from isolated (i.e. far away from CR sources) MCs is most likely dominated by neutral pion decay, and thus such observations can be used to probe CRs with energies above $\approx 280$ MeV, the threshold for pion production in proton-proton interactions.
The emerging picture suggests that the gamma-ray emission from MCs is not affected by CR exclusion, which implies that CRs probe the entire (or a major fraction of the) mass of MCs (see e.g. \cite{CosB,Fermi} for early Cos B and recent Fermi observations, respectively).

Despite that, a detailed discussion of CR exclusion from clouds is still necessary for at least three reasons.
First of all, available gamma-ray observations do not trace CRs of energy smaller than the threshold for pion production. Such CRs play a major role in ionising the MC, and in determining the cloud's dynamics and eventual collapse which leads to star formation (see e.g. \cite{methierry} and references therein).
Second, even when the {\it global} emission from a MC is not affected by CR exclusion, spatial variations of the gamma-ray spectrum within the cloud can still take place, if CRs are excluded from the dense clumps and cores found inside clouds \cite{gab}. Such spatial variations of the cloud spectrum might be detectable by future high angular resolution instruments such as the Cherenkov Telescope Array.
Third, the Fermi observations quoted above refer to clouds characterised by a typical column density of $\approx 10^{22}$ cm$^{-2}$. The giant MC complex in the centre of the Milky Way exhibits much larger column densities, and thus in that particular case CR exclusion might still take place (see discussion below).

%\begin{figure}[t]
%  \centerline{\includegraphics[width=0.5\textwidth]{Fig1-sketch}
%  \includegraphics[width=0.5\textwidth]{Fig3-spectra}}
%  \caption{{\bf Left:}  {\bf Right:} Figures from \cite{morlino}.\label{fig:penetration}}
%\end{figure}

\begin{figure}[t]
%\begin{minipage}{6in}
  \centering
  \raisebox{-0.5\height}{\includegraphics[height=1.48in]{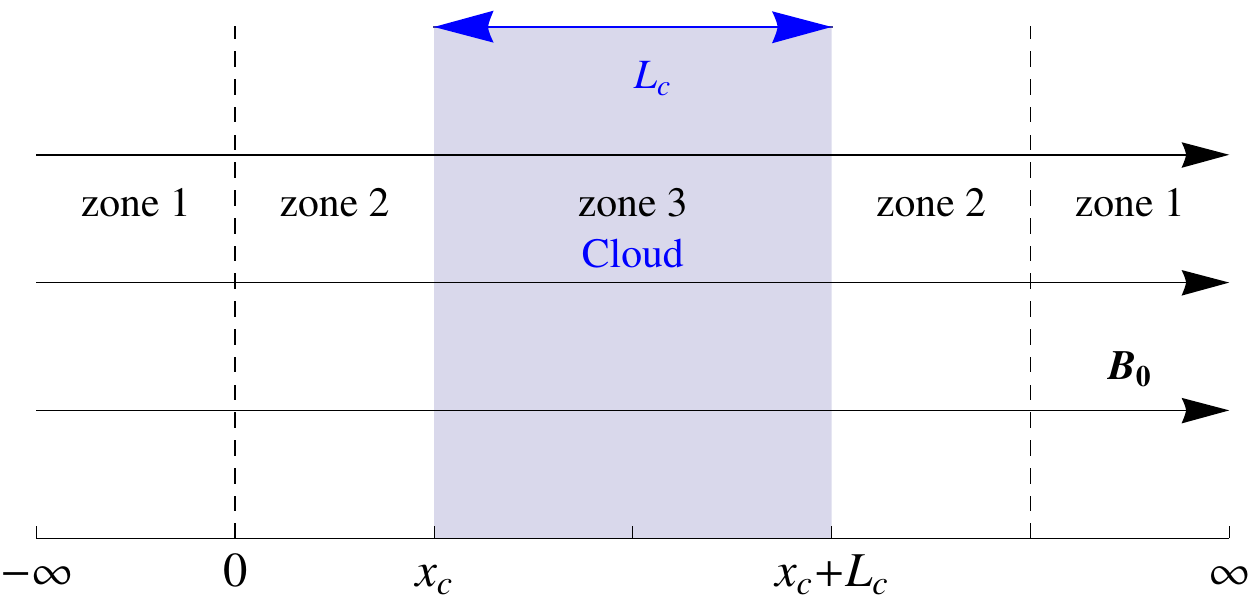}}
  \hspace*{.2in}
  \raisebox{-0.5\height}{\includegraphics[height=1.78in]{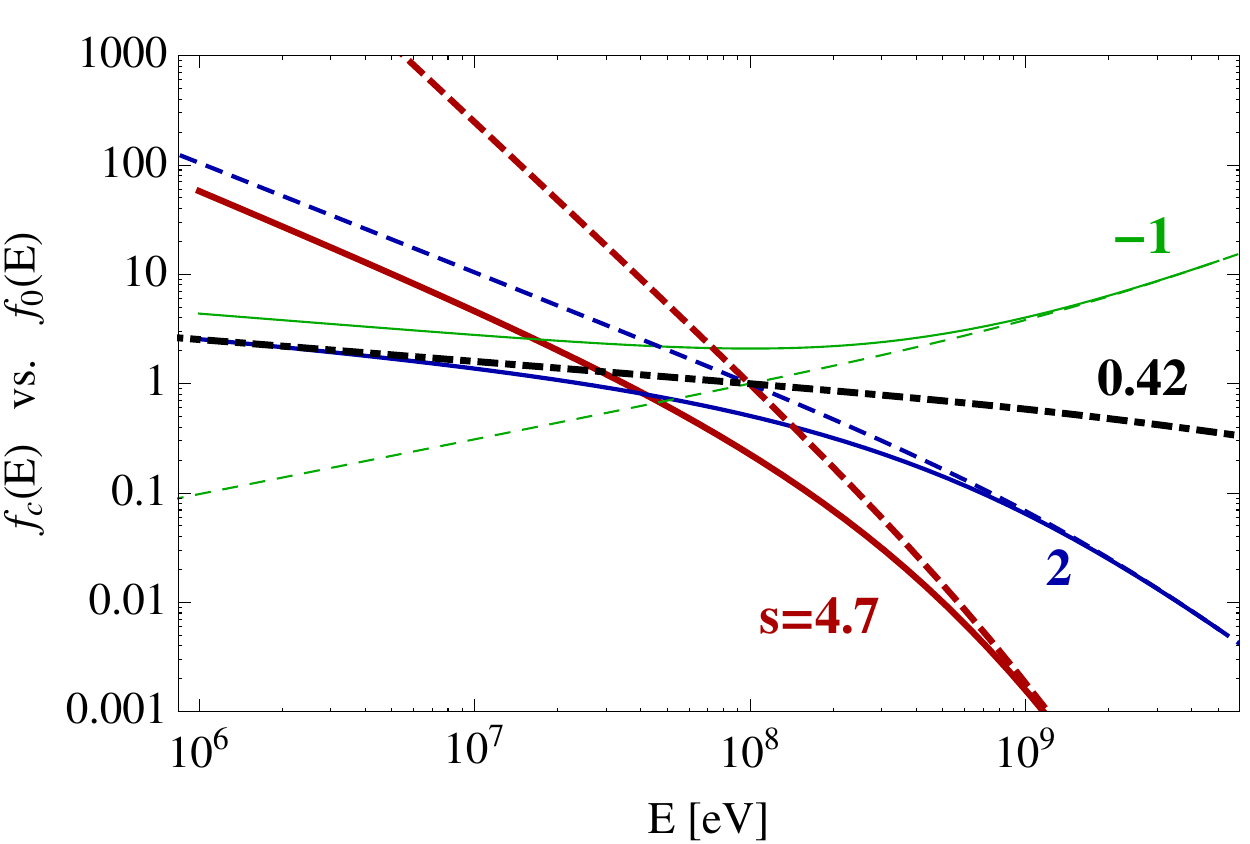}}
%\end{minipage}
\caption{{\bf Left:} Sketch of the MC model considered in the text. {\bf Right:} CR spectrum outside (dashed lines) and inside (solid lines) the MC for different values of the spectral parameter $s$. For $s \sim 0.42$ the two spectra coincide (dot-dashed line). See text for details. Figures from \cite{morlino}.\label{fig:penetration}}
\end{figure}

To describe in a simplified (but still reasonably accurate) way the penetration of CRs into MCs we follow \cite{morlino} and consider the setup represented in the left panel of Figure \ref{fig:penetration}.
A one-dimensional magnetic flux tube in the interstellar medium passes through a MC of size $L_c$ and hydrogen density $n_c$. The magnetic field $B_0$ is assumed to remain spatially constant along the tube, as observed in clouds as long as the gas density does not exceed significantly $\sim 300$ cm$^{-3}$ \cite{crutcher}. A one-dimensional solution can be found over spatial scales smaller or comparable with the coherence length of the interstellar magnetic field, which is of the order of $\lambda_B \approx 50...100$ pc.
In the absence of sources of CRs we can search for a steady state solution of the problem. At distances large enough from the MC (zone 1 in Figure \ref{fig:penetration}), the CR intensity at a given momentum $f_1(p)$ is spatially constant and equal to the galactic CR background. Inside the cloud (zone 3), due to severe energy losses suffered by CRs in the dense gas, one expects $\langle f_3(p) \rangle < f_1(p)$, where the brackets represent a spatial average over zone 3.
The CR intensity in zone 2 can be obtained by solving the one-dimensional and stationary transport equation (e.g. \cite{skilling}):
\begin{equation}
\label{eq:zone2}
v_A ~ \partial_x f_2(p) = D_2 ~ \partial^2_x f_2(p)
\end{equation}
where $x$ is the spatial coordinate along the magnetic field and we assumed the particle diffusion coefficient $D_2$ to be spatially homogeneous. We also neglected the energy loss term which is unimportant in low density environments. The first term in the equation represents an advective flux converging towards the cloud at the Alfven speed $v_A$. This is justified not only in a scenario in which Alfven waves are generated via streaming instability by CRs moving towards the cloud \cite{skillingstrong,cesarsky,morfill}, but also for a generic source of magnetic turbulence. This is because waves are damped in the dense and neutral gas of the MC, and thus in the cloud
vicinity one does not expect any appreciable flux of waves coming from the cloud.

By imposing an equilibrium between CR diffusion and advection one can estimate the size of region 2 to be $x_c \sim D_2/v_A \sim 30 (D_2/10^{27} {\rm cm/s}^2) (v_A/100~{km/s})^{-1}$ pc, which is roughly of the same order of the field coherence length. At this point, Equation \ref{eq:zone2} can be solved in an approximate way after noting that for $0 < x < x_c$ the diffusion term dominates over the advection one. After dropping the advection term one gets: $f_2 = f_1-[f_1-f_2(x_c^-)](x/x_c)$ from which one can compute the flux of CRs entering into the MC as: $2~ [f_2(x_c) v_A - D_2 \partial_x f_2 |_{x_c^-} ] = 2 f_1 v_A$, where the factor of 2 takes into account the fact that the cloud has two sides \cite{skillingstrong,morlino}.
The fact that the entering flux does not depend on the CR diffusion coefficient is a quite remarkable result, and implies a sort of universality of the solution.
Indeed, the spectrum of CRs inside the cloud $\langle f_3(p) \rangle$ can be obtained by balancing the flux of CRs entering the cloud with the rate at which CRs are removed from the cloud due to energy losses:
\begin{equation}
\label{eq:equilibrium}
2 ~ f_1(p) ~ v_A = \frac{L_c}{p^2} \frac{\partial}{\partial p} \left[ \frac{p^3}{\tau_l} \langle f_3(p) \rangle \right]
\end{equation}
where $\tau_l = -p/\dot{p} \propto p^{2.58}/n$ is the momentum loss rate of CRs mainly due to ionisation losses\footnote{The approximate scaling in momentum is valid for particle energies in the range between 100 KeV and 1 GeV \cite{marco}.}.
Clearly, Equation \ref{eq:equilibrium} is valid only when ionisation losses play a role (if they can be neglected one gets the trivial solution $\langle f_3(p) \rangle = f_1(p)$, and dimensional analysis suggests that this indeed happens when:
$
\eta(p) \equiv 2 ~ v_A ~ \tau_l(p)/L_c \le 1
$.
It follows that, for $\eta > 1$ (large momenta) the spectrum of CRs inside the MC is identical to that of galactic CRs, while for $\eta < 1$ (low momenta) the spectrum inside the cloud is suppressed, and a break appear at a particle energy \cite{morlino}:
\begin{equation}
\label{eq:break}
E_{break} \sim 70 ~ \left( \frac{v_A}{100 ~ {\rm km/s}} \right)^{-0.78} \left( \frac{N_H}{3 \times 10^{21} {\rm cm}^{-2}} \right)^{0.78} ~ {\rm MeV}
\end{equation}
which corresponds to $\eta = 1$. In the Equation above, the cloud column density $N_H$ has been normalised to a typical value for MCs in the galactic disk.
The fact that $E_{break}$ is smaller than the energy threshold for neutral pion production in proton-proton interactions implies that the gamma-ray emission from clouds is unaffected. This in turn means that the intensity and spectrum of CRs inside MCs can be deduced straightforwardly from gamma-ray observation, i.e., MCs can be used as CR barometers \cite{sabrinona}.

However, as stated above, the MCs in the galactic centre region constitute a notable exception. In particular, the Sgr B complex encompasses a total mass exceeding $10^7 M_{\odot}$ \cite{ruigi} and exhibits a column density that reaches, in correspondence of the Sgr B2 cloud, a value of few times $10^{24}$ cm$^{-2}$ \cite{andrea}. A naive extrapolation of Equation \ref{eq:break} (which is valid only up to $\sim$ 1 GeV) shows that the break in the CR spectrum inside Sgr B2 has to be located at energies exceeding the GeV. Remarkably, a break in the CR proton spectrum has been observed at a particle energy of $\sim$ 4.5 GeV, following an analysis of Fermi data \cite{ruigi}.
Even though a more accurate modelling of the complex environment of the galactic centre is clearly needed, all thee facts suggests that we might be witnessing the exclusion of low energy CRs from the most massive MCs in our Galaxy.

So far we considered only isolated cloud, located far away from CR sources and thus embedded in the galactic background of CRs. A natural extension of this work would be to study the case of clouds irradiated by a flux of CRs coming from nearby accelerators. Unfortunately, this is not a trivial generalisation of what said so far. First of all, under these conditions the solution of the problem is no longer stationary (the term $\partial_t f$ has to be added on the left side of Equation \ref{eq:zone2}), secondly, the CR intensity could be, in this case, much larger than the typical one in the galactic disk, and thirdly, the penetration of CRs would be asymmetric: if an accelerator is located at the origin of the x axis in Figure \ref{fig:penetration} one would expect to have a net flux of CRs entering into the left side of the MC, and a net flux of CRs going out of the MC at its right side (rather than a converging flow of CRs, as for isolated MCs). 
This problem is very relevant given the fact that many SNR/MC associations have been detected by the Fermi satellite \cite{fermisnrcatalogue} and deserves further attention.

Finally, the exclusion of low energy (MeV) CRs from clouds plays a crucial role in determining the ionisation level of MC, an issue that will be discussed in the next Section.

\section{COSMIC RAY IONISATION OF MOLECULAR CLOUDS}
Tracing CR protons with particle energies below $\sim 280$ MeV, the threshold for neutral pion production, is not straightforward.
Here, we summarise the content of the recent review \cite{methierry} devoted to this issue, where an extended list of references can also be found.

Low energy CRs are the main ionising agents inside MCs, where ionising radiation cannot penetrate, being absorbed due to the large gas column density.
In particular, CRs ionize molecular hydrogen which in turn produces, via an ion-neutral reaction, the pivotal ion $H^+_3$, which plays a central role in the MC chemistry.
In diffuse clouds $H^+_3$ is destroyed by recombination with electrons released by the photoionization
of $C$ into $C^+$ (the most abundant ion in diffuse MCs).
The knowledge of the reaction rates involved in this very simple chemical network allows to infer the CR ionisation rate $\zeta_{CR}$ from the measurement of the $H_3^+$ abundance in a MC.
Under typical MC conditions only the two lowest levels of $H^+_3$ are significantly populated by electrons, and luckily a number of spectral lines originating from the transitions from these
two levels fall in a region of the infrared spectrum observable by ground based infrared telescopes. It has to be stressed that the $H^+_3$ lines are detected in {\it absorption}, and thus an estimate of the $H^+_3$ column density from a given MC can be done solely in the presence of a bright infrared foreground (or embedded) star.

The situation changes dramatically if we consider larger column densities, i.e. dense clouds. There, carbon atoms are mostly in $CO$, and $H^+_3$ is mainly destroyed by the reaction: $H^+_3 + CO \rightarrow HCO^+ +H_2$.
A difficulty with the detection of $H^+_3$ from dense clouds is the fact that for large gas column
densities the background infrared source might be totally obscured by the cloud itself. Thus, alternative
approaches to estimate $\zeta_{CR}$ have been sought. A natural possibility is to search for molecular
lines associated to the molecular ions $HCO^+$ and $DCO^+$ (produced as the result of similar chemical reaction chains). The advantage of this approach resides in the fact that such lines are in emission (no need for a background source!) and fall in the millimeter domain, which is probed by instruments like the 30 meters IRAM telescope. Another advantage of emission lines is that they allow a detailed mapping of extended clouds, a thing which is unfeasible by means of observations of absorption lines, since one would need an unreasonably large number of background sources.

% Figure
\begin{figure}[h]
 \centerline{\includegraphics[width=0.5\textwidth]{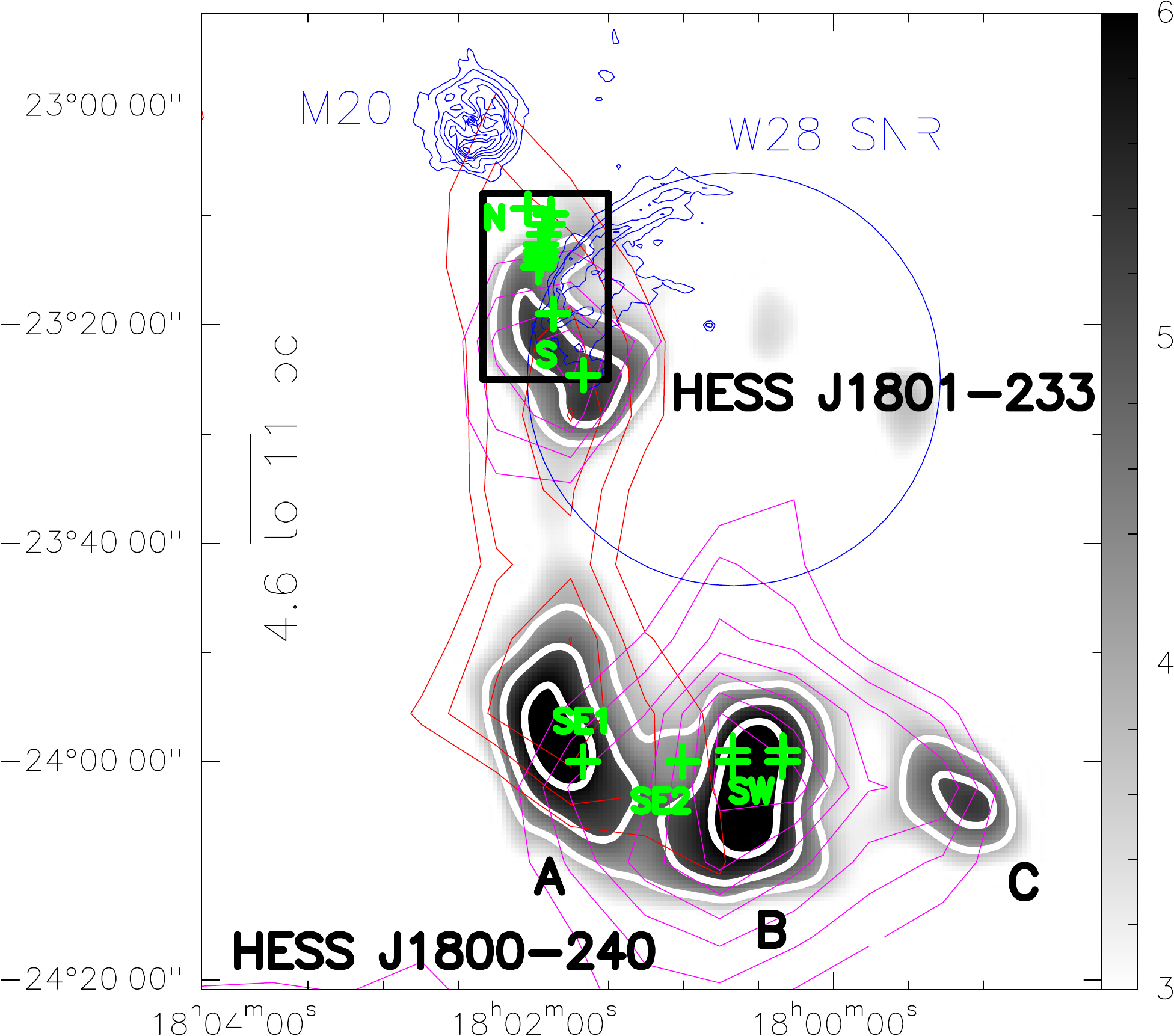}
  \includegraphics[width=0.45\textwidth]{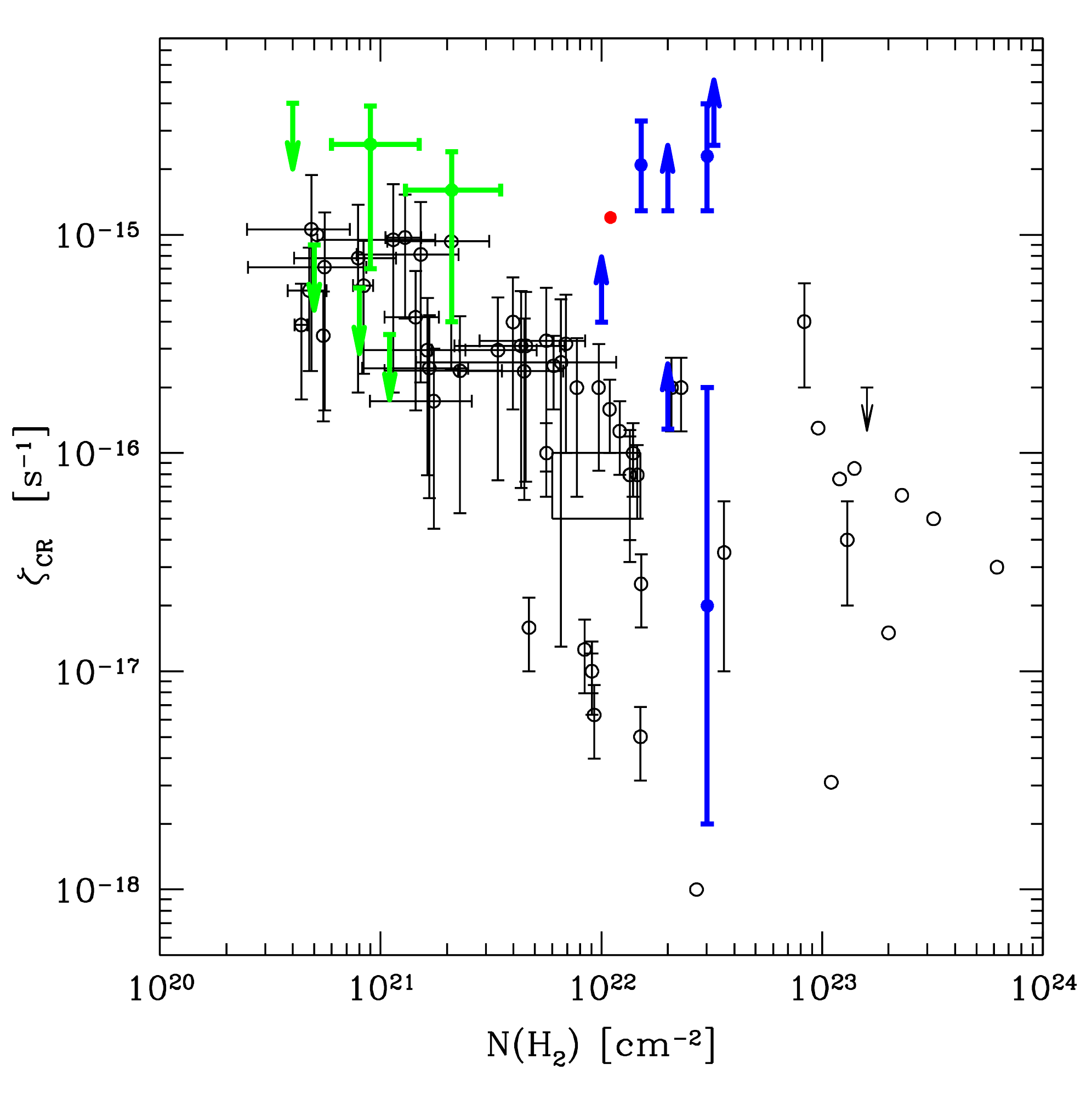}}
  \caption{{\bf Left:} Map of the W28 complex. Greyscale and white contours show the TeV emission seen by H.E.S.S. (levels are 4-6 $\sigma$), red (magenta) contours the CO(1-0) emission integrated over 15-25 km/s (5-15 km/s), and blue contours the 20 cm emission. The blue circle is the approximate boundary of the radio SNR. Green crosses correspond to mm (IRAM) observations aimed at determining the ionisation rate of the molecular material close to the SNR shock \cite{vaupre}. {\bf Right:} CR ionisation rate versus MC column density. Black data points refer to isolated MCs, while colored ones to the SNR/MC associations IC443, W51C, and W28 (green, red, and blue points, respectively) \cite{methierry}.}
\label{fig:mev}
\end{figure}

A compilation of the measurements of $\zeta_{CR}$ in MCs, obtained mainly by means of the two
approaches described above, is given in the right panel of Figure \ref{fig:mev}. Black data points refer to isolated (and gamma-ray quiet) MCs, while colored ones to gamma-ray-bright SNR/MC associations, i.e. MCs spatially associated with SNRs. The presence of gamma-ray emission indicates that the MC material is bombarded by an enhanced (with respect to the CR background) flux of high energy (GeV/TeV) CRs. 
It is clear from Figure \ref{fig:mev} that a trend exists for isolated MCs, namely, the CR ionization rates are a factor of 10...100 larger in diffuse clouds (column densities $N_H \approx 10^{21}
...10^{22}$ cm$^{-2}$) than in dense clouds ($N_H > 10^{22}$ cm$^{-2}$). On the contrary, a different trend is observed for MCs located next to gamma-ray-bright SNRs, for in this case the ionization rates reach large values of the order of $\gtrsim 10^{-15}$ s$^{-1}$, regardless of the cloud column density. Even though this result should be taken with care, being based on the observation of three SNR/MC associations only, it should be nevertheless considered as a first evidence for the fact that MCs belonging to a gamma-ray bright SNR/MC association exhibit, as a general behavior, larger values of the CR ionization rate when compared to isolated clouds. In other words, such MCs are bombarded by a larger flux of low energy (probably $\sim$MeV), ionizing CRs. The associated SNR shock is the most plausible site for the acceleration of suck particles. It is a remarkable fact that the combination of low and high energy observations of SNR/MC associations, and the emerging evidence of a correlation between large gamma-ray fluxes and enhanced ionization rates provide not only additional support to the SNR hypothesis for the origin of Galactic CRs, but also establishes a long sought connection between low and high energy CRs. This will allow us to test models on CR acceleration and propagation over an energy interval of unprecedented breadth, spanning from the MeV to the TeV domain.

\section{CONCLUSIONS AND FUTURE PERSPECTIVES}

In this brief review, we summarised the status of gamma-ray observations of SNRs and MCs and their interpretation, in the context of the SNR hypothesis for the origin of galactic CRs.
Remarkably, we now know that at least some SNRs certainly accelerate CR protons up to multi-GeV energies.
Moreover, the SNR hypothesis passed successfully a number of gamma-ray based tests which have been described above, and provides consistent explanations of many observations of SNRs.
However, we still lack any direct observational evidence for the acceleration of PeV particles at SNR shocks.
Also from a theoretical point of view, the capability of SNRs of accelerating PeV protons at the rate required to explain CR observations is far from being established.
In fact, if, as it seems to be, the transition to extragalactic CRs happens at the energy of the ankle, then the sources of galactic CRs should accelerate protons up to energies well above the PeV domain, a challenge for the standard SNR scenario.

The recent discovery of a proton PeVatron in the galactic centre reminded us that new observations reveal very often unexpected situations.
We now know that at least one source capable of accelerating protons up to PeV energies does exist in our galaxy, and most likely it is not a SNR.
Even though it is not clear whether the CR source located galactic centre can contribute or not to the observed flux of CRs, its very existence suggests that scenarios alternative to the SNR hypothesis for the origin of CRs should not be discarded.

% Acknowledgement
\section{ACKNOWLEDGMENTS}
I would like to thank the organisers of Gamma2016 for their invitation, and F. Aharonian, C. Ceccarelli, P. Cristofari, A. Goldwurm, P. Hennebelle, J. Krause, A. Marcowith, M. Menu, T. Montmerle, G. Morlino, L. Nava, M. Padovani, E. Parizot, V. Ptuskin, and R. Terrier for collaborations and discussions.

% References

\nocite{*}
\bibliographystyle{aipnum-cp}%
\bibliography{sample}%

\end{document}